\documentclass[conference]{IEEEtran}

\usepackage{amsmath,amssymb,amsthm,mathrsfs}
\usepackage{epsfig,epsf,subfigure,graphicx,graphics}
\usepackage{url}

\IEEEoverridecommandlockouts

\newtheorem{lemma}{Lemma}
\newtheorem{theorem}{Theorem}

\newcommand{\SNR}{\mathsf{SNR}}
\newcommand{\INR}{\mathsf{INR}}
\newcommand{\C}{\mathsf{C}^\mathsf{B}}
\newcommand{\E}{\mathrm{E}}

\newcommand{\sym}{\mathrm{sym}}

\newcommand{\lp}{\left(}
\newcommand{\rp}{\right)}

\newcommand{\lbp}{\left\{}
\newcommand{\rbp}{\right\}}

\newcommand{\mcal}{\mathcal}
\newcommand{\mscr}{\mathscr}
\newcommand{\what}{\widehat}

\graphicspath{{fig/}}

\title{Interference Mitigation Through Limited Receiver Cooperation: Symmetric Case}

\author{
\authorblockN{I-Hsiang Wang and David N. C. Tse}
\authorblockA{Wireless Foundations\\
University of California at Berkeley,\\
Berkeley, California 94720, USA\\
\textsf{\{ihsiang, dtse\}@eecs.berkeley.edu}}
\thanks{This work was supported by National Science Foundation under grant \# CCF-0830796.}
}

\begin{document}
\maketitle

\begin{abstract}
Interference is a major issue that limits the performance in wireless networks, and cooperation among receivers can help mitigate interference by forming distributed MIMO systems. The rate at which receivers cooperate, however, is limited in most scenarios. How much interference can one bit of receiver cooperation mitigate? In this paper, we study the two-user Gaussian interference channel with conferencing decoders to answer this question in a simple setting. We characterize the fundamental gain from cooperation: at high $\SNR$, when $\INR$ is below 50\% of $\SNR$ in dB scale, one-bit cooperation per direction buys roughly one-bit gain per user until full receiver cooperation performance is reached, while when $\INR$ is between 67\% and 200\% of $\SNR$ in dB scale, one-bit cooperation per direction buys roughly half-bit gain per user. The conclusion is drawn based on the approximate characterization of the symmetric capacity in the symmetric set-up. We propose strategies achieving the symmetric capacity universally to within 3 bits. The strategy consists of two parts: (1) the transmission scheme, where superposition encoding with a simple power split is employed, and (2) the cooperative protocol, where {quantize-binning} is used for relaying.
\end{abstract}


\section{Introduction}
In modern communication systems, interference is one of the fundamental factors that limit performance: a receiver is only interested in retrieving information from its own transmitter, while the information-carrying signals become interference to other users due to the broadcast and superposition nature of wireless channels. The simplest information theoretic model for studying this issue is the \emph{interference channel}. Characterizing the capacity region is a long-standing open problem, except for several special cases. Recently Etkin, Tse, and Wang characterize the capacity region of the Gaussian interference channel to within one bit \cite{EtkinTse_07} by using a superposition coding scheme with a simple power-split configuration and by providing new upper bounds.

In the above interference channel set-up, transmitters or receivers are not allowed to communicate with one another, and hence each user has to combat interference on its own. In various applications, however, nodes are not isolated, and transmitters/receivers can exchange certain amount of information. Since the nodes are distributed due to the physical constraints, the amount of information they can exchange is limited. Therefore, one of the fundamental questions is, how much {\it interference} can limited {\it transmitter/receiver cooperation} mitigate?

In this paper, we consider a two-user Gaussian interference channel with {\it conferencing decoders} to answer this question regarding receiver cooperation. Conferencing among encoders/decoders has been studied in \cite{Willems_83}, \cite{BrossLapidoth_08}, \cite{MaricYates_07}, \cite{DaboraServetto_06}, \cite{Simeone_08}, and \cite{YuZhou_08}. Our model is similar to those in \cite{Simeone_08} and \cite{YuZhou_08} but in an interference channel set-up. The work in \cite{Simeone_08} characterizes the capacity region of the compund MAC with unidirectional conferencing between decoders and provides achievable rates for general set-up but is not able to establish a constant-gap result. The work in \cite{YuZhou_08} considers one-sided Gaussian interference channels with unidirectional conferencing between decoders and characterizes the capacity region in strong interference regimes and the asymptotic sum capacity at high $\SNR$. For general receiver cooperation, works including \cite{Host-Madsen_06} and \cite{PrabhakaranViswanath_09}, investigate cooperation in interference channels with a set-up where the cooperative links are of the same band as the links in the interference channel. In particular, \cite{PrabhakaranViswanath_09} characterizes the sum capacity of Gaussian interference channels with in-band receiver cooperation to within 40 bits. Our work, on the other hand, is focused on the Gaussian interference channel with orthogonal receiver cooperation.

We propose a strategy achieving the symmetric capacity universally to within 3 bits in the symmetric set-up, regardless of channel parameters. The three-bit gap is the worst-case gap which can be loose in some regimes, and it is vanishingly small at high $\SNR$ when compared to the capacity.The strategy consists of two parts: (1) the transmission scheme, describing how transmitters encode their messages, and (2) the cooperative protocol, describing how receivers exchange information and decode messages. For transmission, both transmitters use superposition coding \cite{HanKobayashi_81} with a simple power-split configuration, which is the same as that in the case without cooperation \cite{EtkinTse_07}, to encode messages. For the cooperative protocol, it is appealing to apply the decode-forward or compress-forward schemes, originally proposed in \cite{CoverElGamal_79} for the relay channel,  like most works dealing with more complicated networks, including \cite{DaboraServetto_06}, \cite{Simeone_08}, \cite{YuZhou_08}, \cite{Host-Madsen_06}, \cite{KramerGastpar_05}, etc. It turns out neither compress-forward nor decode-forward achieves capacity to within a constant number of bits universally for the problem at hand. On the other hand, \cite{CoverKim_07}, \cite{Kim_07}, and \cite{AvestimehrDiggavi_09} observe that the conventional compress-forward scheme \cite{CoverElGamal_79} may be improved by the destination directly decoding the sender's message instead of requiring to first decode the quantized signal of the relay. We use such an improved compress-forward scheme as our cooperative protocol. Each receiver first quantizes its received signal at an appropriate distortion, bins the quantization codeword and sends the bin index to the other receiver.
Each receiver then decodes its own information based on its own received signal and the received bin index. It turns out that this simple "one-round" cooperative protocol is sufficient to achieve within a constant gap to the symmetric capacity in the symmetric case. In the general asymmetric channel, it turns out that this simple protocol is not sufficient while a more sophisticated "two-round" protocol is \cite{WangTse_09}.

\section{Problem Formulation}\label{sec_Formulation}
\begin{figure}[htbp]
{\center
\includegraphics[width=2.5in]{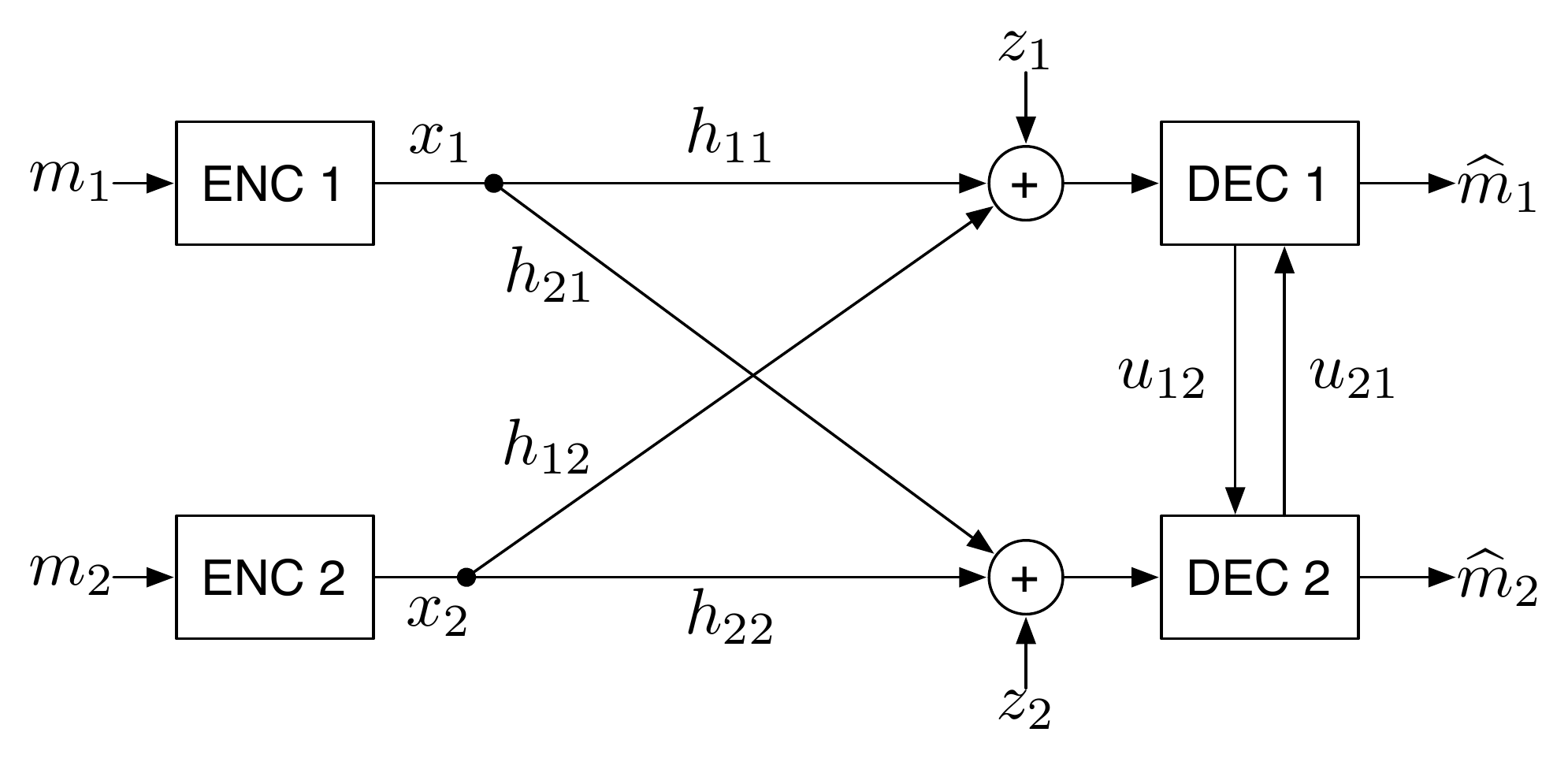}
\caption{Channel Model}
\label{fig_ChModel}
}
\end{figure}

The Gaussian interference channel with conferencing decoders is depicted in Fig. \ref{fig_ChModel}.

\subsection{Channel Model}
\subsubsection{Transmitter-Receiver Links}
The transmitter-receiver links are modeled as the \emph{normalized} Gaussian interference channel:
\begin{align*}
y_1 &= h_{11}x_1 + h_{12}x_2 + z_1\\
y_2 &= h_{21}x_1 + h_{22}x_2 + z_2,
\end{align*}
where the additive noise processes $\{z_i[n]\}$, ($i=1,2$), are independent $\mathcal{CN}(0,1)$, i.i.d. over time. In this paper, we use $[.]$ to denote time indices. Transmitter $i$ intends to convey message $m_i$ to receiver $i$ by encoding it into a block codeword $\{x_i[n]\}_{n=1}^N$, with transmit power constraints
\begin{align*} 
\frac{1}{N}\sum_{n=1}^N\E\Big[ \big\lvert x_i[n] \big\lvert^2 \Big] \le 1,\ i=1,2,
\end{align*}
for arbitrary block length $N$. Note that outcome of the encoder depends solely on its own message. Messages $m_1,m_2$ are independent.
Define channel parameters
\begin{align*}
\SNR_i := |h_{ii}|^2,\ \INR_i := |h_{ij}|^2,\ i,j=1,2,\ i\ne j.
\end{align*}
%

\subsubsection{Receiver-Cooperative Links}
The receiver-cooperative links are noiseless with finite capacity $\C_{ij}$ from receiver $i$ to $j$. Encoding must satisfy causality constraints: for any time index $n=1,2,\ldots, N$, $u_{21}[n]$ is only a function of $\{y_2[1],\ldots, y_2[n-1], u_{12}[1], \ldots, u_{12}[n-1]\}$, and $u_{12}[n]$ is only a function of $\{y_1[1],\ldots, y_1[n-1], u_{21}[1], \ldots, u_{21}[n-1]\}$.


\section{Achievable Strategy and Symmetric Capacity to within 3 Bits}\label{sec_Achieve}
We focus on the symmetric set-up, namely, $\SNR = \SNR_1 = \SNR_2$, $\INR = \INR_1 = \INR_2$, and $\C = \C_{12} = \C_{21}$.

For the symmetric set-up, a natural performance measure is the symmetric capacity 
$C_{\sym} := \sup \left\{ R: (R,R) \in \mscr{C}\right\}$, where $\mscr{C}$ is the capacity region.

\subsection{Outline of the Strategy}
In our model, note that arbitrarily large number of rounds\footnote[1]{By multiple rounds we mean that one receiver can decide what to send to the other receiver after it receives the side information from the other, and so on so forth.} are allowed for conferencing among receivers. Remarkably, with the proposed strategy, one-round\footnote[2]{One-round means that each receiver decides on its own what to send to the other receiver.} conference is sufficient to achieve $C_{\sym}$ to within constant number of bits universally.

The strategy proposed in this section consists of three basic ingredients: {\it superposition coding} at transmitters, {\it quantize-binning} for relaying, and a decoder keeping track of the codebook structure when figuring out quantization codewords. Due to space constraint, we give an outline.

{\flushleft \it Superposition coding}:\par
For each transmitter, it splits its own message into common and private (sub-)messages. Each common message is aimed at both receivers, while each private one is aimed at its own receiver. Each message is encoded into a Gaussian random codeword with certain power. As \cite{EtkinTse_07} points out, since the private signal is undesired at the unintended receiver, a reasonable configuration is to make the private interference at or below the noise level so that it does not cause much damage and can still convey additional information in the direct link if it is stronger than the cross link. When the interference is stronger than the desired signal, simply set the whole message to be common. 

{\flushleft \it Quantize-binning}:\par
Upon receiving its signal from the transmitter-receiver link, each receiver does not decode messages immediately. Instead, each receiver, serving as a relay, first quantizes its signal by a pre-generated Gaussian quantization codebook with certain distortion, and then sends out a bin index determined by a pre-generated binning function. How should we set the distortion? Note that both its own private signal and the noise it encounters are not of interest to the other receiver. Therefore, a natural configuration is to set the distortion level equal to {\it the maximum of noise power and private signal power level}.

{\flushleft \it Decoding}:\par
After retrieving the receiver-cooperative side information, that is, the bin index, the receiver decodes the two common messages and its own private message, by searching in transmitters' codebooks for a codeword triple (indexed by the two common messages and the user's own private message) that is jointly typical with its received signal and some quantization point (codeword) in the given bin. If there is no such unique codeword triple, it declares an error.

\subsection{Comparison with the Conventional Compress-Forward}


Note that the main difference between our cooperative protocol and the conventional compress-forward with Gaussian vector quantization lies in the \emph{decoding} procedure and the chosen distortion. In the conventional Gaussian compress-forward, the decoder first searches in the bin for one quantization codeword that is jointly typical with its received signal from its own transmitter \emph{only}, assuming that the two received signals are jointly Gaussian. This may not be true since a single user may not transmit at the capacity in its own link, which results in ``holes" in signal space. As a consequence, this scheme may not utilize the dependency of two received signals well and cause larger distortions. Our scheme, on the other hand, utilizes the dependency in a better way by \emph{jointly} deciding the quantization codeword and the message triple, consequently allows smaller distortions, and is able to reveal the beneficial side information to the other receiver.

\begin{figure}[htbp]
{\center
\subfigure[Optimal Scheme]{\includegraphics[width=1.7in]{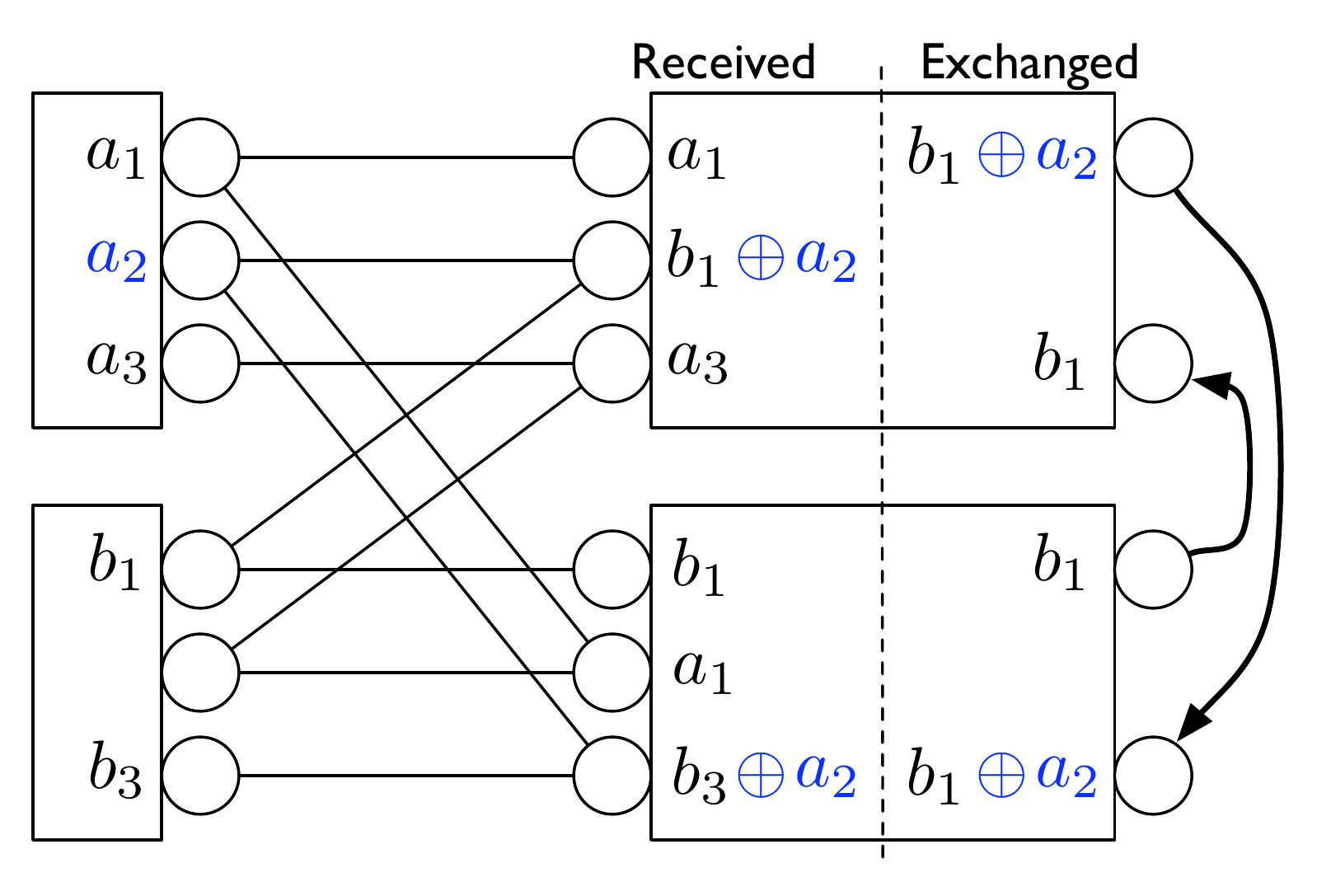}}
\subfigure[Compress-Forward]{\includegraphics[width=1.7in]{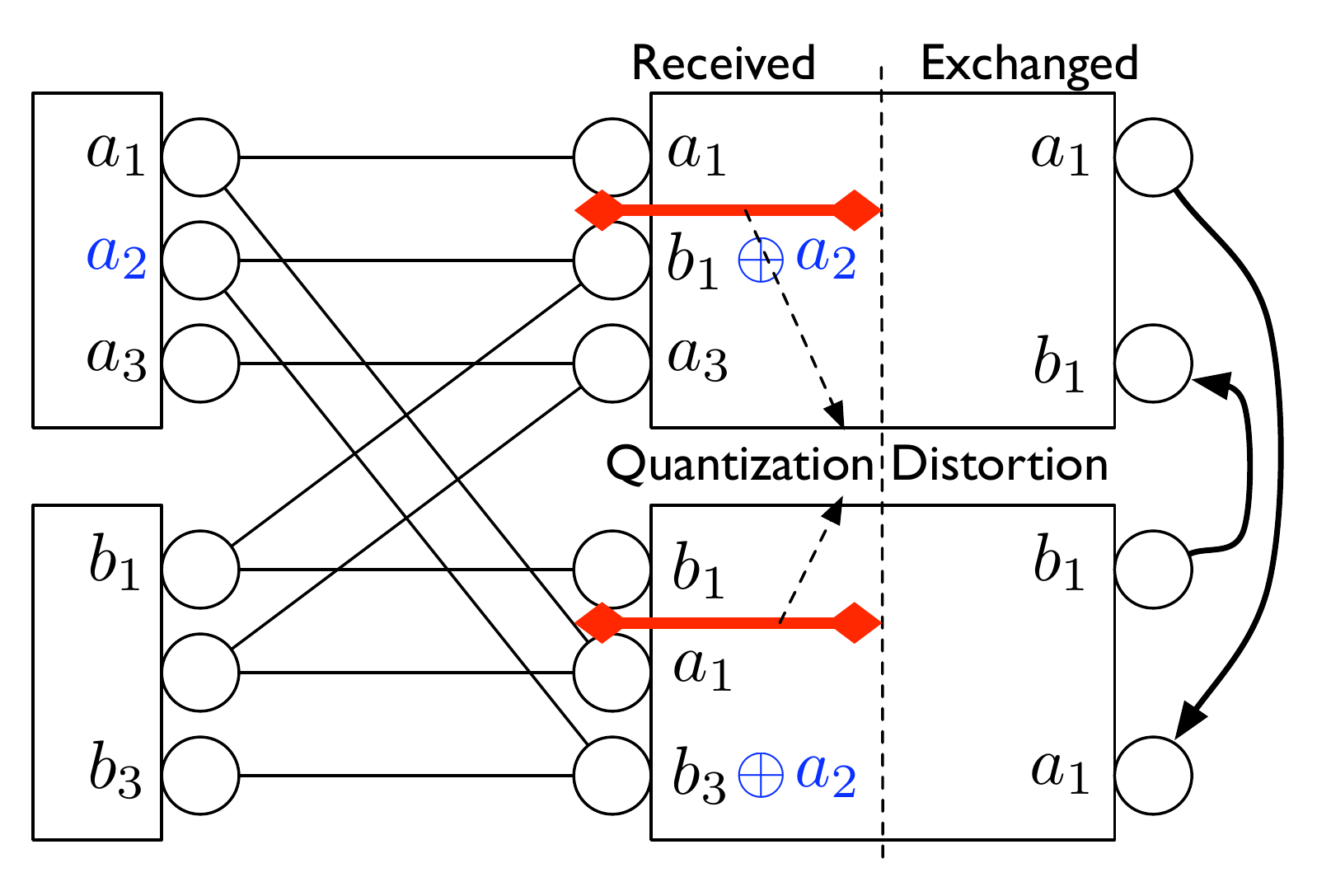}}
\caption{An Example Channel}
\label{fig_Examples}
}
\end{figure}

We give an example to illustrate the above observations. In this example channel, set $\INR$ to be 2/3 of $\SNR$ in dB scale, that is, $\log\INR=\frac{2}{3}\log\SNR$. Besides, set $\C=\frac{1}{3}\log\SNR$. To better convey the key ideas, we make use of the \emph{linear deterministic channel} (LDC) proposed in \cite{AvestimehrDiggavi_09}.  The corresponding channel is depicted in Fig. \ref{fig_Examples}. The bits $\lp a_1,a_2,a_3\rp$ and $\lp b_1,0,b_3\rp$ can be viewed as the binary expansions of the transmitted signals. Note that in this example, one bit in the LDC corresponds to $\frac{1}{3}\log\SNR$ in the Gaussian channel. As a baseline, without cooperation the optimal sum rate is 4 bits in the LDC. With one-bit cooperation in each direction in the LDC, the optimal sum rate is 5 bits. The scheme is depicted in Fig. \ref{fig_Examples}.(a).

From its corresponding LDC, one can see that the two received signals of the Gaussian channel, $\lp y_1,y_2\rp$, are not jointly Gaussian. The reason is that, suppose they are jointly Gaussian, the conditional distribution of $y_2$ given $y_1$ should be marginally Gaussian. As Fig. \ref{fig_Examples} suggests, however, conditioning on receiver 1's signal results in a hole at the second level of receiver 2's signal, which was occupied by $a_1$. Therefore, transmitter 2's common codebook is not dense enough to make the conditional distribution of $y_2$ given $y_1$ marginally Gaussian.

Now, suppose compress-forward assuming joint Gaussianity of the received signals is used for receivers to cooperate. This is a standard approach to evaluate achievable rates of Gaussian channels using compress-forward in the literature. The incorrect assumption results in larger quantization distortions, as depicted in Fig. \ref{fig_Examples}.(b). The information sent from receiver 1 to receiver 2, $a_1$, is \emph{redundant}, and cannot help mitigate interference $a_2$. Hence, the achievable sum rate is 4 bits (3 bits for user 1 and 1 bit for user 2), which is the same as that without cooperation and is one bit less than the optimal performance. Recall that 1 bit in the LDC corresponds to $\frac{1}{3}\log\SNR$ in the Gaussian channel, therefore the performance loss is unbounded as $\SNR\rightarrow\infty$.


Our scheme is very similar to \emph{extended hash-and-forward} proposed in \cite{Kim_07}, in which it is pointed out that the scheme has no advantage over the conventional compress-forward in a single-source single-relay setting. Due to the above mentioned issues, however, we recognize in our problem where the channel consists of two source-destination pairs and two relays, the scheme has an unbounded advantage over the conventional compress-forward in certain regimes.

\subsection{Achievable Symmetric Rate}\label{subsec_AchRate}
Due to space constraint, we give the following coding theorem without proof. Let $R_{ic}$ and $R_{ip}$ denote the rates for user $i$'s common message and private message respectively, for $i=1,2$.
\begin{theorem}
The rate tuple $\lp R_{1c},R_{2c},R_{1p},R_{2p}\rp$ satisfying the following constraints are achievable:\\
\underline{Constraints at receiver 1}:
\begin{align*}
R_{1p} &\le I\lp x_{1};y_1|x_{1c},x_{2c}\rp + \lp\C_{21} - \xi_1\rp^+\\
R_{2c}+R_{1p} &\le I\lp x_{2c},x_{1};y_1|x_{1c}\rp + \lp\C_{21} - \xi_1\rp^+ \\
R_{1c}+R_{1p} &\le I\lp x_{1};y_1|x_{2c}\rp + \lp\C_{21} - \xi_1\rp^+ \\
R_{1c}+R_{2c}+R_{1p} &\le I\lp x_{1},x_{2c};y_1 \rp + \lp\C_{21} - \xi_1\rp^+,
\end{align*}
where $\xi_1 = I\lp \what{y}_2; y_2| x_{1c},x_1,x_{2c},y_1\rp$, and $x_{ic}$ is the common codebook generating random variable for $i=1,2$.
\begin{align*}
R_{1p} &\le I\lp x_{1};y_1,\what{y}_2 | x_{1c},x_{2c}\rp\\
R_{2c}+R_{1p} &\le I\lp x_{2c},x_{1};y_1,\what{y}_2 | x_{1c}\rp\\
R_{1c}+R_{1p} &\le I\lp x_{1};y_1,\what{y}_2 | x_{2c}\rp\\
R_{1c}+R_{2c}+R_{1p} &\le I\lp x_{1},x_{2c};y_1,\what{y}_2 \rp.
\end{align*}
where $\what{y}_2\overset{d}{=}y_2+\what{z}_2$ is the quantization codebook generating random variable, and $\what{z}_2\sim\mcal{CN}\lp0.\Delta_2\rp$, independent of everything else. $\Delta_2$ is the quantization distortion at receiver 2.
{\flushleft \underline{Constraints at receiver 2}: the above constraints with index ``1" and ``2" exchanged.}
\end{theorem}

In the rest of this paper, we focus on the symmetric set-up. Besides, for simplicity we assume the typical case where $\SNR>1$ and $\INR > 1$. 
We defer the full treatment of general asymmetric set-up in a subsequent paper \cite{WangTse_09}.

\begin{lemma}[Achievable Symmetric Rate]\label{lem_Achievable}
When $\SNR \le \INR$,
\begin{align*}
&R_{\sym} =  
\min\lbp \begin{subarray}{l}
\log\lp 1+\SNR \rp +(\C -1)^+,\log\lp 1+\SNR+\INR\rp-1,\\
\frac{1}{2}\big[  \log \lp 1+\SNR+\INR \rp + (\C-1)^+ \big],\\
\frac{1}{2}\big[ \log\lp1+2\SNR+2\INR+|h_{11}h_{22}-h_{12}h_{21}|^2 \rp -1\big]
\end{subarray}\rbp
\end{align*}
is achievable. When $\SNR > \INR$, 
\begin{align*}
&R_{\sym} = \\ 
&\min\lbp \begin{subarray}{l}
\log\lp 1+\frac{\SNR}{\INR}+\INR \rp + \lp\C - \log3\rp^+ -1,\log\lp1+\SNR \rp -2,\\
\frac{1}{2}\big[ \log\lp1+\SNR+\INR\rp + \log\lp2+\frac{\SNR}{\INR}\rp + \lp\C - \log3\rp^+ -2 \big],\\
\frac{1}{2}\big[ \log\lp1+2\SNR+2\INR+|h_{11}h_{22}-h_{12}h_{21}|^2 \rp -3\big]
\end{subarray}\rbp
\end{align*}
is achievable.
\end{lemma}

Next, 
we have outer bounds for symmetric capacity:
\begin{lemma}[Outer Bounds for Symmetric Capacity]\label{thm_SymOutBd}
$C_{\sym} \le \overline{C}_{\sym}$ with 
\begin{align*}
\overline{C}_{\sym} = \min\lbp \begin{subarray}{l}
\log(1+\SNR)+ \min\left\{ \C,\log\left(1+\frac{\INR}{1+\SNR}\right) \right\},\\
\log\left(1+\INR+\frac{\SNR}{1+\INR}\right) + \C,\\
\frac{1}{2}\log\left(1+\SNR+\INR\right) + \frac{1}{2}\log\left(1+\frac{\SNR}{1+\INR}\right) + \frac{1}{2}\C,\\
\frac{1}{2}\log\lp 1+2\SNR+2\INR + |h_{11}h_{22} - h_{12}h_{21}|^2 \rp
\end{subarray}\rbp.
\end{align*}

\end{lemma}


With Lemma \ref{lem_Achievable} and \ref{thm_SymOutBd}, we establish the following theorem:
\begin{theorem}[Constant Gap to Symmetric Capacity]\label{thm_ConstGap}
The strategy can achieve the symmtric capacity to within 3 bits; namely,
\begin{align*}
R_{\sym} \le C_{\sym} \le \overline{C}_{\sym} \le R_{\sym} + 3
\end{align*}
\end{theorem}

 



%


\section{Generalized Degrees of Freedom}\label{sec_Dof}
Generalized degrees of freedom (g.d.o.f.) characterization, originally proposed in \cite{EtkinTse_07}, is an asymptotic capacity characterization in high $\SNR$ regime. For our problem, it is appealing to define a similar notion for characterizing the high-$\SNR$ asymptotic performance, in the following way: let
$\lim_{\SNR\rightarrow\infty}\frac{\log\INR}{\log\SNR} = \alpha$, $\lim_{\SNR\rightarrow\infty}\frac{\C}{\log\SNR} = \kappa$, 
and define the number of generalized degrees of freedom per user as
\begin{align*}\label{eq_Dof}
\tag{\ref{sec_Dof}-$*$}d\lp \alpha,\kappa\rp := \lim_{\begin{subarray}{c} \mathrm{fix}\ \alpha,\kappa\\ \SNR\rightarrow\infty\end{subarray}}\frac{C_{\sym}}{\log\SNR},
\end{align*}
if the limit exists. With fixed $\alpha$ and $\kappa$, however, there are certain channel realizations under which \eqref{eq_Dof} has different values and hence the limit does not exist. This happens when $\alpha = 1$, where the phases of the channel gains matter both in inner and outer bounds. In particular, its value can depend on whether the system MIMO matrix is well-conditioned or not.



Instead of claiming that the limit (\ref{eq_Dof}) exists for {\it all} channel realizations, we pose a reasonable distribution, namely, i.i.d. uniform distribution, on the phases, show that the limit exists {\it almost surely}, and define the limit to be the number of {\it generalized degrees of freedom} per user. Details are omitted here due to space constraint. In particular, \eqref{eq_Dof} is the same as the symmetric capacity normalized by the interference-free capacity per user in the corresponding linear deterministic channel (LDC) except for $\alpha=1$.


Now that the number of g.d.o.f. is well-defined, we can give the following theorem:
\begin{theorem}[Number of Generalized Degrees of Freedom Per User]\label{thm_Dof}
We have a direct consequence from Lemma \ref{thm_SymOutBd} and Theorem \ref{thm_ConstGap}:
\begin{align*}
d = \lbp 
\begin{array}{ll}
\min\lbp 1, \max\lp \alpha,1-\alpha\rp + \kappa, 1-\frac{\alpha-\kappa}{2} \rbp, &0 \le \alpha < 1\\
\min\lbp \alpha, 1 + \kappa, \frac{\alpha+\kappa}{2} \rbp, &\alpha \ge 1
\end{array}
\right. 
\end{align*}
\end{theorem}

\begin{figure}[htbp]
{\center
\includegraphics[width=2in]{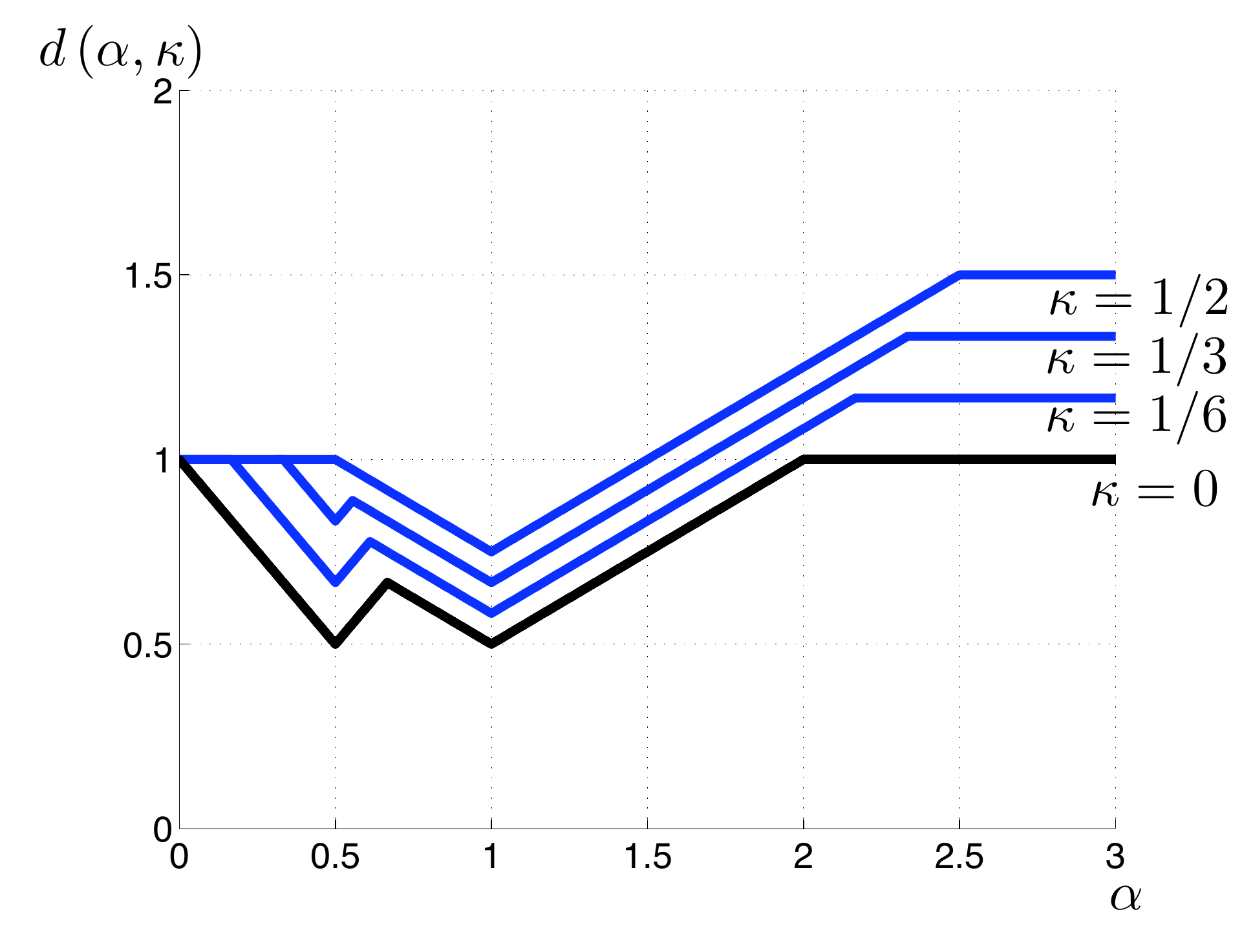}
\caption{Generalized Degrees of Freedom}
\label{fig_Dof}
}
\end{figure}

Numerical plots for g.d.o.f. are given in Fig. \ref{fig_Dof}. We observe that at different values of $\alpha$, the gain from cooperation varies. By investigating the g.d.o.f., we conclude that at high $\SNR$, when $\INR$ is below 50\% of $\SNR$ in dB scale, one-bit cooperation per direction buys roughly one-bit gain per user until full receiver cooperation performance is reached, while when $\INR$ is between 67\% and 200\% of $\SNR$ in dB scale, one-bit cooperation per direction buys roughly half-bit gain per user until saturation.

The fundamental behavior of the gain from receiver cooperation is explained in the rest of this section, by looking at two particular points: $\alpha=\frac{1}{2}$ and $\alpha=\frac{2}{3}$. Furthermore, we use the linear deterministic channel (LDC) for illustration.
\begin{figure}[htbp]
{\center
\subfigure[]{\includegraphics[width=1.7in]{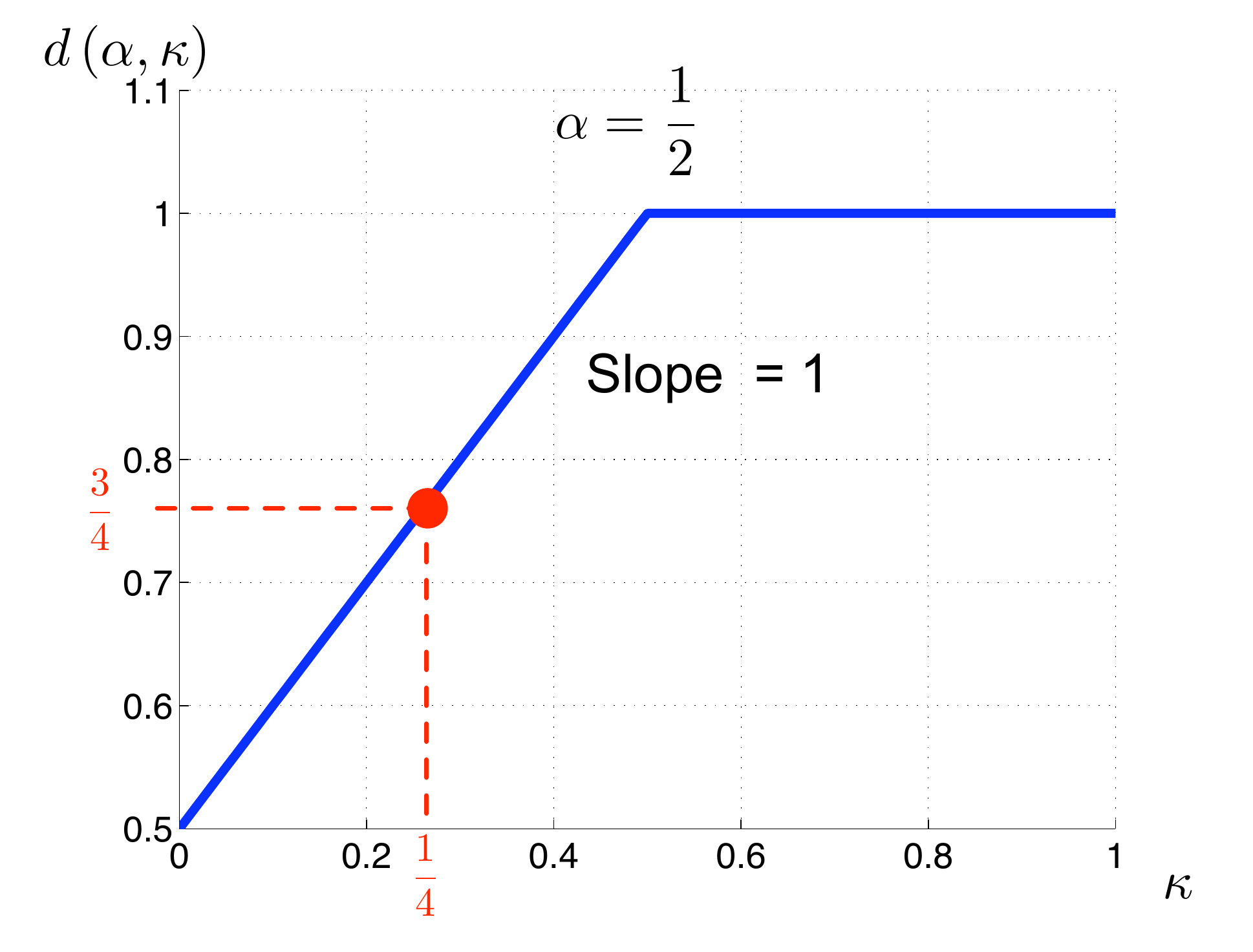}}
\subfigure[]{\includegraphics[width=1.7in]{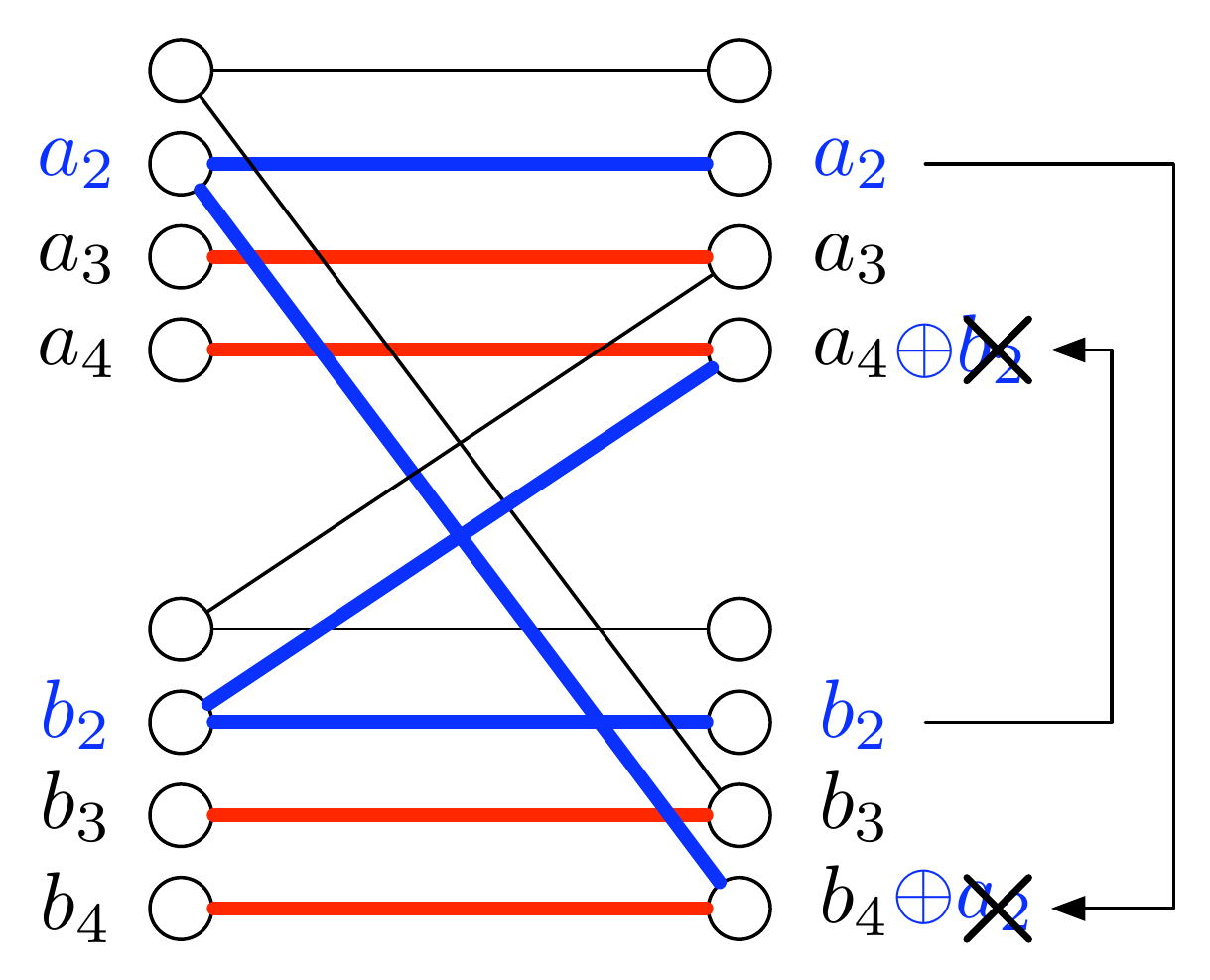}}
\subfigure[]{\includegraphics[width=1.7in]{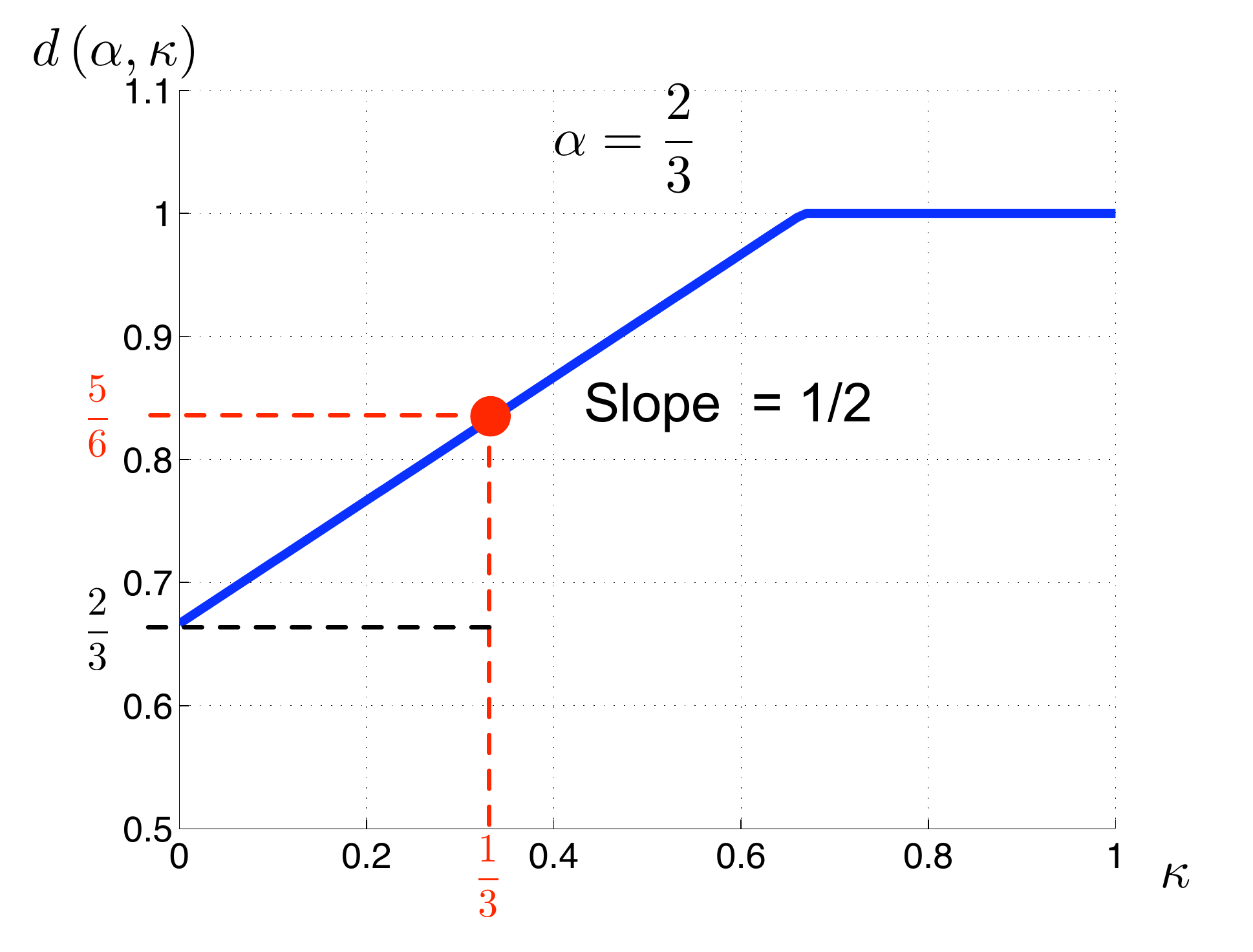}}
\subfigure[]{\includegraphics[width=1.7in]{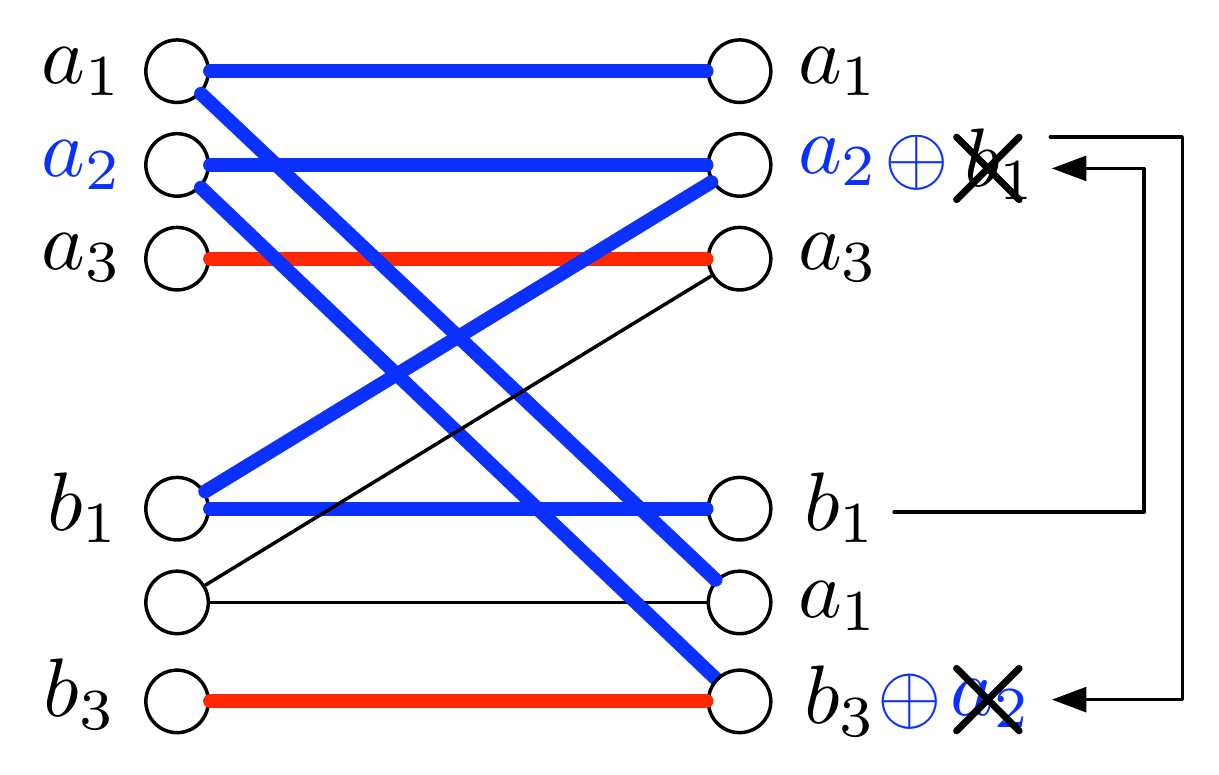}}
\caption{Gain from Cooperation}
\label{fig_Regimes}
}
\end{figure}

At $\alpha = \frac{1}{2}$, the plot of $d$ versus $\kappa$ is given in Fig. \ref{fig_Regimes}.(a). The slope is 1 until full receiver cooperation performance is reached, implying that one-bit cooperation buys one more bit per user. We look at a particular point $\kappa=\frac{1}{4}$ and use its corresponding LDC (Fig. \ref{fig_Regimes}.(b)) to provide insights. Note that 1 bit in the LDC corresponds to $\frac{1}{4}\log\SNR$ in the Gaussian channel, and since $\C \approx \frac{1}{4}\log\SNR$, in the corresponding LDC each receiver is able to sent one-bit information to the other. Without cooperation, the optimal way is to turn on bits not causing interference, that is, the \emph{private} bits $a_3,a_4,b_3,b_4$. We cannot turn on more bits without cooperation since it causes collisions, for example, at the fourth level of receiver 2 if we turn on $a_2$ bit. Now with receiver cooperation, we want to support two more bits $a_2,b_2$. Note that prior to turning on $a_2,b_2$, there are ``holes" left in receiver signal spaces, and turning on each of these bits only causes one collision at one receiver. Therefore, we need 1 bit in each direction to resolve the collision at each receiver. We can achieve 3 bits per user in the corresponding LDC and $d = \frac{3}{4}$ in the Gaussian channel. We cannot turn on more bits in the LDC since it causes collisions while no cooperation capability is left.

At $\alpha = \frac{2}{3}$, the plot of $d$ versus $\kappa$ is given in Fig. \ref{fig_Regimes}.(c). The slope is $\frac{1}{2}$ until full receiver cooperation performance is reached, implying that two-bit cooperation buys one more bit per user. We look at a particular point $\kappa=\frac{1}{3}$ and use its corresponding LDC (Fig. \ref{fig_Regimes}.(d)) to provide insights. Note that now 1 bit in the LDC corresponds to $\frac{1}{3}\log\SNR$ in the Gaussian channel, and since $\C \approx \frac{1}{3}\log\SNR$, in the corresponding LDC each receiver is able to sent one-bit information to the other. Without cooperation, the optimal way is to turn on bits $a_1,a_3,b_1,b_3$. We cannot turn on more bits without cooperation since it causes collisions, for example, at the second level of receiver 2 if we turn on $a_2$ bit. Now with receiver cooperation, we want to support one more bit $a_2$. Note that prior to turning on $a_2$, there are no ``holes" left in receiver signal spaces, and turning on $a_2$ causes collisions at \emph{both} receivers. Therefore, we need 2 bits in total to resolve collisions at both receivers. We can achieve 5 bits in total in the corresponding LDC and $d = \frac{5}{6}$ in the Gaussian channel. We cannot turn on more bits in the LDC since it causes collision while no cooperation capability is left.

From above examples and illustrations, we see that whether \emph{one cooperation bit buys one more bit} or \emph{two cooperation bits buy one more bit} depends on whether there are ``holes" in receiver signal spaces before increasing data rates. The ``holes" play a central role not only in why the conventional compress-forward is suboptimal in certain regimes, as mentioned in the previous section, but also in the fundamental behavior of the gain from receiver cooperation. We notice that in \cite{PrabhakaranViswanath_09}, there is a similar behavior about the gain from cooperation as discussed in Section 3.2. of \cite{PrabhakaranViswanath_09}. We conjecture that the behavior can be explained via the concept of ``holes" as well.

\section{General Asymmetric Case: An Example}\label{sec_Conclusions}

\begin{figure}[hbtp]
{\center
\subfigure[Suboptimal One-round Scheme]{\includegraphics[width=1.7in]{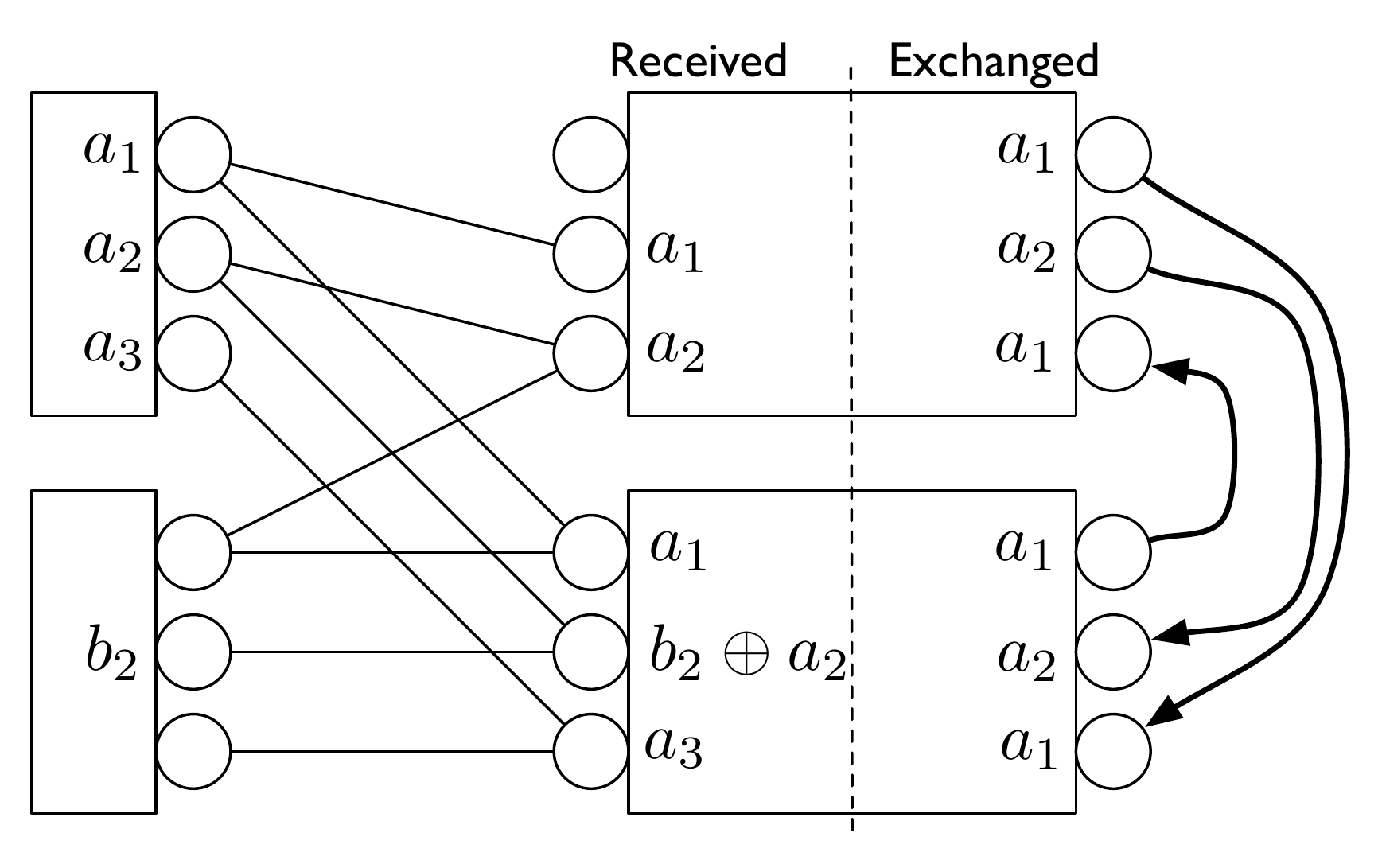}}
\subfigure[Optimal Two-round Scheme]{\includegraphics[width=1.7in]{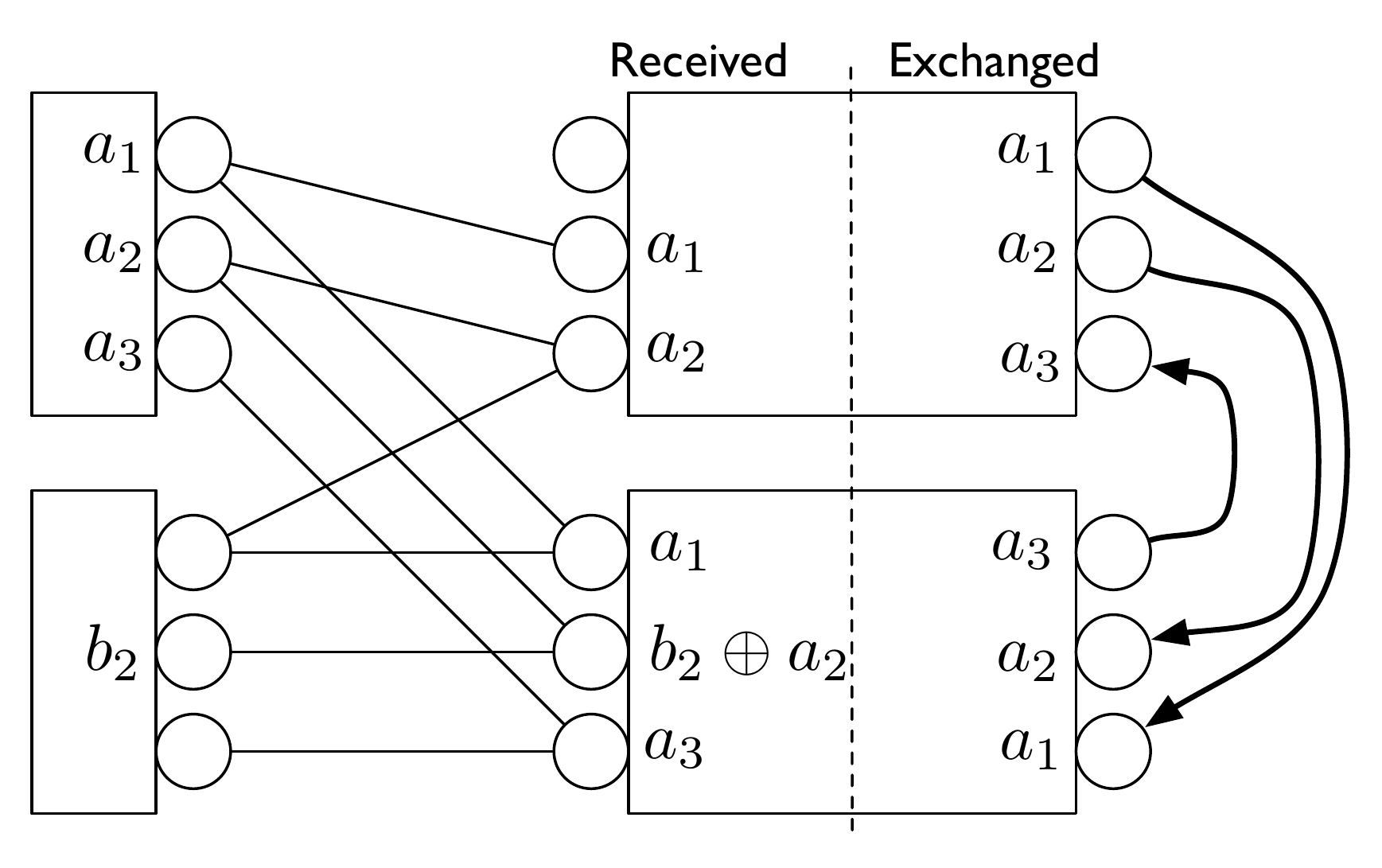}}
\caption{A Motivating Example for the Two-Round Scheme}
\label{fig_Asymm}
}
\end{figure}

In this paper, we propose a one-round scheme achieving the symmetric capacity to within 3 bits. The proposed one-round scheme, however, is not sufficient to achieve the capacity region to within a constant number of bits in general. As an example, consider the LDC in Fig. \ref{fig_Asymm}. If receiver 2 quantizes at it private signal level, it can only forward $a_1$ to receiver 1 and achieves $R_1$ up to 2 bits. On the other hand, if receiver 2 first decodes $b_2,a_3$ and then forwards $a_3$ to receiver 1, it achieves $R_1=3$ bits. In \cite{WangTse_09}, this problem with general asymmetric set-up is investigated. We implement a two-round strategy and show that it can achieve the capacity region universally to within 2 bits per user. 

\bibliographystyle{ieeetr}

\end{document}